\documentclass[final,5p,times,twocolumn]{elsarticle}

\usepackage{graphicx}
\usepackage{dcolumn}
\usepackage{bm}
\usepackage{amsmath}

\begin{document}

\title{Non-adiabatic pumping of single electrons affected by magnetic fields}

\author[ptb]{C. Leicht}
\ead{christoph.leicht@ptb.de}
\author[ptb]{B. Kaestner}

\address[ptb]{Physikalisch-Technische Bundesanstalt, Bundesallee 100, 38116 Braunschweig, Germany}

\author[lat1,lat2]{V. Kashcheyevs}

\address[lat1]{Institute for Solid State Physics, University of Latvia, Riga LV-1063, Latvia}
\address[lat2]{Faculty of Physics and Mathematics, University of Latvia, Ze{\c{l}\c{l}}ustreet 8, Riga LV-1002, Latvia}

\author[ptb]{P. Mirovsky}
\author[ptb]{T. Weimann}
\author[ptb]{K. Pierz}
\author[ptb]{H. W. Schumacher}


\date{\today}

\begin{abstract}
Non-adiabatic pumping of discrete charges, realized by a dynamical quantum dot in an AlGaAs/GaAs heterostructure, is studied under influence of a perpendicular magnetic field. Application of an oscillating voltage in the GHz-range to one of two top gates, crossing a narrow wire and confining a quantum dot, leads to quantized pumped current plateaus in the gate characteristics. The regime of pumping one single electron is traced back to the diverse tunneling processes into and out-of the dot. Extending the theory to multiple electrons allows to investigate conveniently the pumping characteristics in an applied magnetic field. In this way, a qualitatively different behavior between pumping even or odd numbers of electrons is extracted.
\end{abstract}

\journal{Physica E}
\maketitle

\section{Introduction}

Well defined pumping of discrete charges using an excitation frequency $f$ has attracted much interest in an effort to realize a quantum standard for the ampere~\cite{averin1986,mills2006}. Amongst others, integrated single-electron circuits~\cite{nishiguchi2006} as well as quantum information processing~\cite{barnes2000} have been proposed. Many approaches to obtain such a quantized current have been tested, following the pioneering work two decades ago~\cite{geerligs1990,kouwenhoven1991,pothier1992}. In all attempts the acid test for a well defined current is the current plateau - as a function of some controlling voltage applied at a gate - at current values of $I=nef$. $n$~is the integer number of electrons with elementary charge $e$ pumped at frequency $f$.

Combining the need for a high current value and high accuracy remains a big challenge. Therefore, devices suitable for parallelization and hence multiplication of the current output have been put forward in two recent approaches. Beside the turnstile device introduced by Pekola $et$ $al.$~\cite{pekola2008}, implemented in metallic systems with fixed tunnel barriers, another approach uses tunable tunnel barriers in a semiconductor nanowire~\cite{blumenthal2007}, operated by a single voltage parameter~\cite{kaestner2008,fujiwara2008,kaestner2008b,maire2008}.
In recent works on AlGaAs based devices of the latter type it was shown that a perpendicular applied magnetic field significantly modifies the region where quantization appears as well as the slope of the quanzited plateaus~\cite{kaestner2009,wright2008}. The origin of this effect is still not understood, but could potentially be used to improve the accuracy be several orders of magnitude~\cite{kaestner2009}.

Here we offer a first step towards quantification of this dependence by applying a recently proposed model for dynamical quantum dot (QD) initialization~\cite{kashcheyevs2009}. This model links the plateaus shape to decay rate ratios between different charging states of the QD and therefore connects it to microscopic properties.
To this end we will first analyze a relevant set of rate equations for the complete parameter range and then find a region in which the above model is valid. In the second part, the model is used to extract an even-odd characteristic of the plateaus as a perpendicular magnetic field is applied.

\section{Experimental Details}

\begin{figure}
\centering
\includegraphics[scale=0.8]{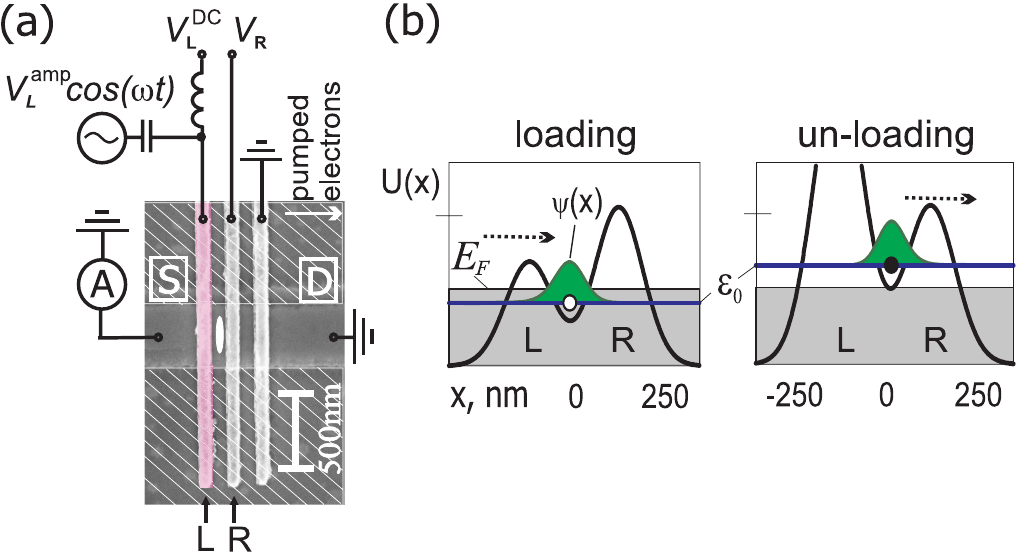}
\caption{\label{fig:device} (Color online) SEM picture of the device shown in (a). Gate voltages are indicated, showing gate L colored in red as being modulated. (S) and (D) mark soure and drain. The hatched regions are depleted of the 2D electron gas, defining a wire. A quasi-bound state is formed between gates L and R, indicated by the white ellipse. The third gate is not used and grounded. (b) Schematic of the potential along the channel during loading (left) and unloading (right) of the quasibound state $\Psi (x)$.}
\end{figure}

A picture of one of the two devices used for our measurements is shown in Fig.~\ref{fig:device}(a). Three metallic top gates, 100 nm in width, are crossing a wire etched in a $n$-type AlGaAs heterostructure. The distance between the gate centers is 250 nm and the bottom gate is not used and grounded. Two devices (labeled 1 and 2) differ in channel width, $w_1=800$ nm, $w_2=900$ nm, and carrier density ${n_{e,1}}=2.1\times 10^{11}$ cm$^{-2}$ and ${n_{e,2}}=2.8\times 10^{11}$ cm$^{-2}$. All measurements were performed in a $^3$He/$^4$He dilution cryostat at temperatures below 50 mK. For quantized charge pumping the devices have to be operated in the following way: Applying sufficiently large negative voltages to gate L and R forms a QD between them. An additional sinusodial modulation of frequency $f$ is then coupled to gate L. If the amplitude is high enough, the energy of the lowest quasi-bound state $\epsilon_0$ in the QD is pushed below the Fermi energy of the leads $E_\text{F}$  in the first half of the cycle and can be loaded with electrons from source. During the second half of the cycle the left barrier is raised fast enough to prevent the electrons from tunneling back to source. At the same time $\epsilon_0$ is raised above $E_\text{F}$, so the electrons can be unloaded to drain. This results in a current created without an external bias voltage as sketched in Fig.\ref{fig:device}(b).

\section{Quantized pumping regime}

To realize transport of a single electron during one oscillation cycle, it is necessary to keep the tunnel couplings of the barriers in a regime, where exactly one electron is loaded into the QD from source and subsequently unloaded to drain. First we will define a quantization region were this condition is fulfilled~\cite{zole2009}.

The voltages creating the barriers L and R are represented by dimensionless parameters $V_L$ and $V_R$, respectively. $V_L(t)=V^\text{DC}_L + V^\text{amp}_L \cos(\omega t)$ is composed of a DC offset voltage and a radio frequency signal with amplitude $V^\text{amp}_L$. It can be assumed that the tunnel couplings $\Gamma^l (V_L, V_R)$ and $\Gamma^r (V_L, V_R)$ of the left and right barrier respectively depend exponentially on the closing potential barrier height~\cite{buettiker1990}. In Ref.~\cite{kaestner2008} it is shown that the resulting time dependence, not only of  log $\Gamma^l$ but also of log $\Gamma^r$ can be well approximated by a cosine function, taking into account that tunneling through one barrier is additionally affected by the other barrier. Hence we assume in dimensionless units: \begin{align} \label{eq:tunnelcouplings}
  \Gamma^l ( V_L, V_R) = \exp [V_L (t) - \gamma V_R] , \\
  \Gamma^r ( V_L, V_R) = \exp [V_R - \gamma V_L (t)] ,
\end{align}
where $0 < \gamma < 1$ is the cross talk ratio.
The value of $\epsilon_0$ is determined by both voltages in a symmetric way:
\begin{equation}
 \epsilon_0 (t) = - (V_L (t) + V_R).
 \label{eq:groundstate}
\end{equation}
Without loss of generality we can set $f = 1$ in $\omega \equiv 2 \pi f$. The relative position of the $\epsilon_0$ with respect to the Fermi energy is fixed by the condition
\begin{equation}
  \epsilon_0 + \xi_0 = 0.
  \label{eq:energyposition}
\end{equation}
Here $\xi_0$ is a dimensionless parameter which represents the Fermi energy in the leads.
Thus the shape of the quantization steps is completely fixed by the parameters $\gamma$, $\xi_0$ and $V^\text{amp}_L$. Regarding the temperature $T$ the following assumptions are made:
\begin{enumerate}[i)]
  \item $T$ is much smaller than the energy modulation amplitude. This allows us to replace the full Fermi function by a step function.
  \item $T$ is much larger than the tunneling rates and the modulation frequency. This allows us to use a simple rate equation.~\cite{kaestner2008} 
\end{enumerate}
The level crossing condition~(\ref{eq:energyposition}) is satisfied in the time moments $t = t_0 + m$ and $t = - t_0 + m$, where $m$ is an integer and
\begin{equation}
  t_0 = \frac{1}{2\pi}\text{arccos}\frac{\xi_0-V_R-V^\text{DC}_L}{V^\text{amp}_L}.
  \label{eq:time}
\end{equation}
The range of ($V^\text{DC}_L$, $V_R$) at which $t_0$ is real (for fixed $V^\text{amp}_L$) determines the region of parameters where the current generation is possible within the model. During one period from $t = -t_0$ till $t = 1 - t_0$ the energy level is below the Fermi surface for $-t_0 < t < +t_0$ and above it for $+t_0 < t < 1 - t_0$.
Corresponding rate equations can be formulated as:
\begin{align}
   &\frac{\partial P}{\partial t} = \left( \Gamma^l + \Gamma^r \right)[1-P(t)], \ -t_0<t<+t_0 ,\\
   &\text{tunneling into the QD is possible, out is prohibited.}\nonumber \\ 
   &\frac{\partial P}{\partial t} = - \left( \Gamma^l + \Gamma^r \right)[P(t)] \text{,  for}+t_0<t<1-t_0 ,\\
   &\text{tunneling into the QD is prohibited, out is possible.}\nonumber
\end{align}
 
Imposing a periodic boundary condition in time makes the periodic stationary solution $P(t)$ unique. The instantaneous current flowing into the QD from the lead $\alpha = l,r$ is calculated as
\begin{equation}
  I^\alpha=\frac{\Gamma^\alpha(t)}{\Gamma^l(l)+\Gamma^r(t)} \ \frac{\partial P(t)}{\partial t}
\end{equation}
Corresponding to this model current measurements in lead~$\alpha$ would ideally give
\begin{equation}
  I_{DC}= \int_0^1 I^\alpha (t) \ dt
\end{equation}
 
\begin{figure}
  \centering
  \includegraphics[scale=0.43]{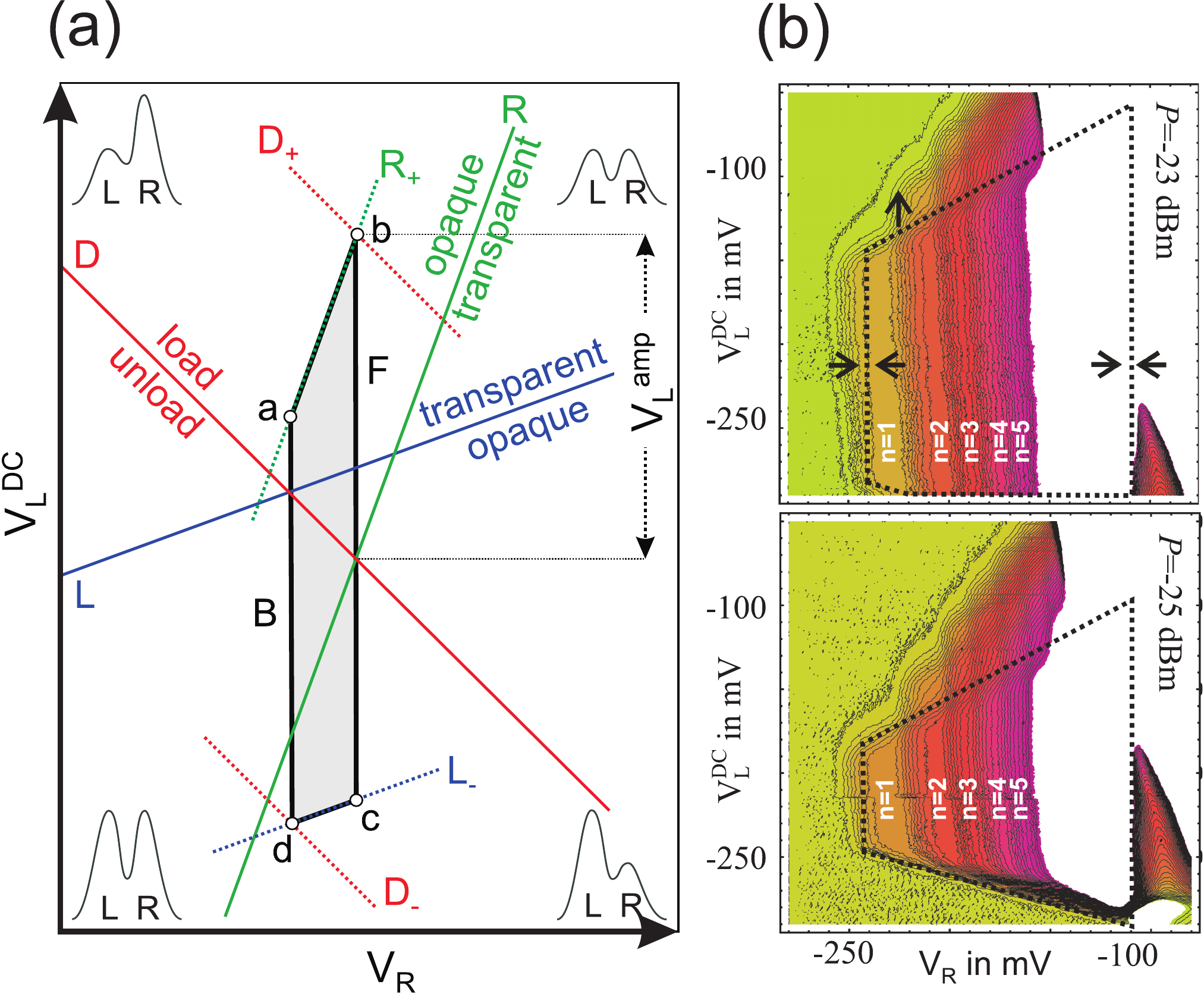}
  \caption{\label{fig:2dmodel} (Color online) (a) The region of $V_{R}$ and $V^\text{DC}_{L}$ where efficient transportation of electrons from source to drain is possible is shaded in gray. The derivation and explanation are given in the text. (b) Typical characteristic of pumped current of device 1 measured as a function of $V^\text{DC}_L$ and $V_R$ at $f=500$~MHz, $P=-25$~dBm and $P=-23$~dBm. While the lower and upper line shift with increasing power, the left and right line stay fixed.}
\end{figure}

We will now discuss the parameter range of $V_R$ and $V^\text{DC}_L$ where efficient transportation of electrons is possible, shown in Fig.~\ref{fig:2dmodel}(a)~\cite{zole2009}. As sketched by the four pictograms in the corners, the heights of the left and right barrier decrease with increasing the corresponding applied voltage to the top gates. 
The blue line L indicates $\Gamma^l = 1$ i.e. the transition between the left barrier being transparent (above, $\Gamma^l \gg 1$) or opaque (below $\Gamma^l \ll 1$) for charge transfer within one period $1/f$ . In a similar manner, the green line indicates $\Gamma^r=1$ i.e. the transition of the right barrier from transparent (below, $\Gamma^r \gg 1$) to opaque (above, $\Gamma^r \ll 1$). The tilt of the lines relative to the axis is determined by the cross talk factor $\gamma \neq 0$. 
 While the lines L and R are related to the tunnel couplings the red line D is related to the position of $\epsilon_0$ relative to the  Fermi energy, see Eq.~\ref{eq:energyposition}. At line D $\epsilon_0$ crosses the Fermi energy ($\epsilon_0 = - \xi_0$). Above this line the level is below the Fermi energy and thus the QD is loaded if tunnel coupling to at least one lead is sufficient large ($\Gamma^{\alpha} \gg 1$). In contrast, below line D it becomes energetically favourable to unload the QD if it is occupied and sufficiently coupled.
 
Note that until now, we have considered only the DC part of the applied voltage. 
In Fig.~\ref{fig:2dmodel} the rf-modulation of the left barrier is taken into account by a parallel vertical shift of the three lines in both directions. The magnitude of the shift is $V^\text{amp}_L$, as indicated by an arrow on the right. The two vertically shifted D-lines are marked  $\text{D}_+$ and $\text{D}_-$. 
Let us first consider the loading of the QD from the source. Due to the additional AC component loading the QD is possible even for $V^\text{DC}_L$ and $V_R$ voltages down to $\text{D}_-$ during one part of the cycle. For quantized current pumping the loading must occur from the source i.e. via the left barrier. Therefore the crossing point of the $\text{D}_-$ line and the $\text{L}_-$ line marks one extremal point (d) of the pumping region. Only above $\text{D}_-$ and above $\text{L}_-$ loading from the source is possible.  The upper boundary of the pumping parameter region can be found when considering unloading the QD to the drain. Therefore the line $\text{D}_+$ must be considered. Only below $\text{D}_+$ unloading of the QD is energetically allowed. Additionally for unloading to the right drain the right barrier has to be transparent. Thus the other extremal point of the pumping regime is marked by (b) i.e. the crossing point of $\text{D}_+$ and $\text{R}_+$. Again, only below $\text{D}_+$ and $\text{R}_+$ unloading to the drain is possible. 
Connecting the extremal points and the lines $\text{L}_-$, $\text{R}_+$ by vertical lines defines the AC pumping region (abcd). 

Qualitatively, this region is bordered as follows: At step B the QD has been completely loaded exclusively from source during $-t_0<t<t_0$. The current outside the pumping region drops because the QD unloads to source again instead to drain during $t_0<t<1-t_0$. Outside step R$_+$ the current drops because of unsufficient unloading to drain during $t_0<t<1-t_0$. Outside step F loading during $-t_0<t<t_0$ takes place predominantly from the drain instead from source. Since the electron is also emitted to drain during $t_0<t<1-t_0$ the average dc current drops to zero. Outside step L$_-$ the QD is unsufficiently loaded through the source barrier, leading again to a current drop.

In Fig.~\ref{fig:2dmodel}(b) measurements of device 1 are plotted for different rf-amplitudes. The size of the pumping region (marked by dotted lines) increases along the $V^\text{DC}_L$-axis with increasing rf-amplitude, but the position of the left and right line remain fixed. This effect agrees with the experimental results, hence all four lines B, F, L$_-$ and R$_+$ can be identified. The different slope at the L$_-$ step might result from neglecting excited states. Lowering the right barrier and consequently $\epsilon_0$ may open additional transport channels through the left barrier into excited states. In contrast to the experimental data where several quantized plateaus appear, only one electron is considered in this model. To describe the multi-electron case, for simplicity only the transition region across line B will be considered next.

The transition regions of L$_-$ and R$_+$ can in principle be arbitrarily well separated from those of B and F by choosing a sufficiently large modulation amplitude. In particular for the measurements considered below, a range of $V^{\text{DC}}_L$ has been identified, where the transition across line B is not significantly affected by that of L, R and F. Therefore, the relevant processes for deviation from the quantized current value are tunneling events of the previously loaded electron back to source.
Whenever the electron remains on the QD it will be emitted to drain during $t_0<t<1-t_0$, so we only have to focus on $-t_0<t<t_0$. At this time $\Gamma^r \approx 0$ can be assumed. Hence, we can simplify our problem to a QD, which decays to source only and becomes increasingly isolated. This picture can be generalized to describe more than one electron, such that the final charge state determines the average number of electrons being transfered.

Such a dynamical QD has already been theoretically investigated in Ref.~\cite{kashcheyevs2009}: A decay cascade model is proposed in which the confining potential is characterized by a fixed sequence of decay rate ratios $\Gamma^l_n/\Gamma^l_{n-1}=e^{\delta_n}$, one rate per each QD occupation number $n$. The rates $\Gamma^l_n$ are assumed to decrease uniformly and exponentially in time $t$ and control gate voltage, corresponding to $V_R$ above. This leads to an analytic solution for the occupation-number probability distribution which is parameterized by a set of $\delta_n$'s. An explicit fitting formula for the average number of captured electrons has been derived for $\delta_n \gg 1$:

\begin{equation}
  \langle n \rangle = \sum^N_{k=1}e^{-e^{-\alpha V_R + \Delta_k}} \quad \text{where} \quad \Delta_k = \sum^k_{i=1} \delta_i.
  \label{eq:fit}
\end{equation}
Thus every plateau is described in length and slope by a significant $\delta$, whereas  $\delta_n$ relates to the ($n-1$)th plateau.
Note that apart from the dimension $\alpha$ is identical to the cross talk ratio $\gamma$.

\section{Experimental results}

\begin{figure}
  \centering
  \includegraphics[scale=0.31]{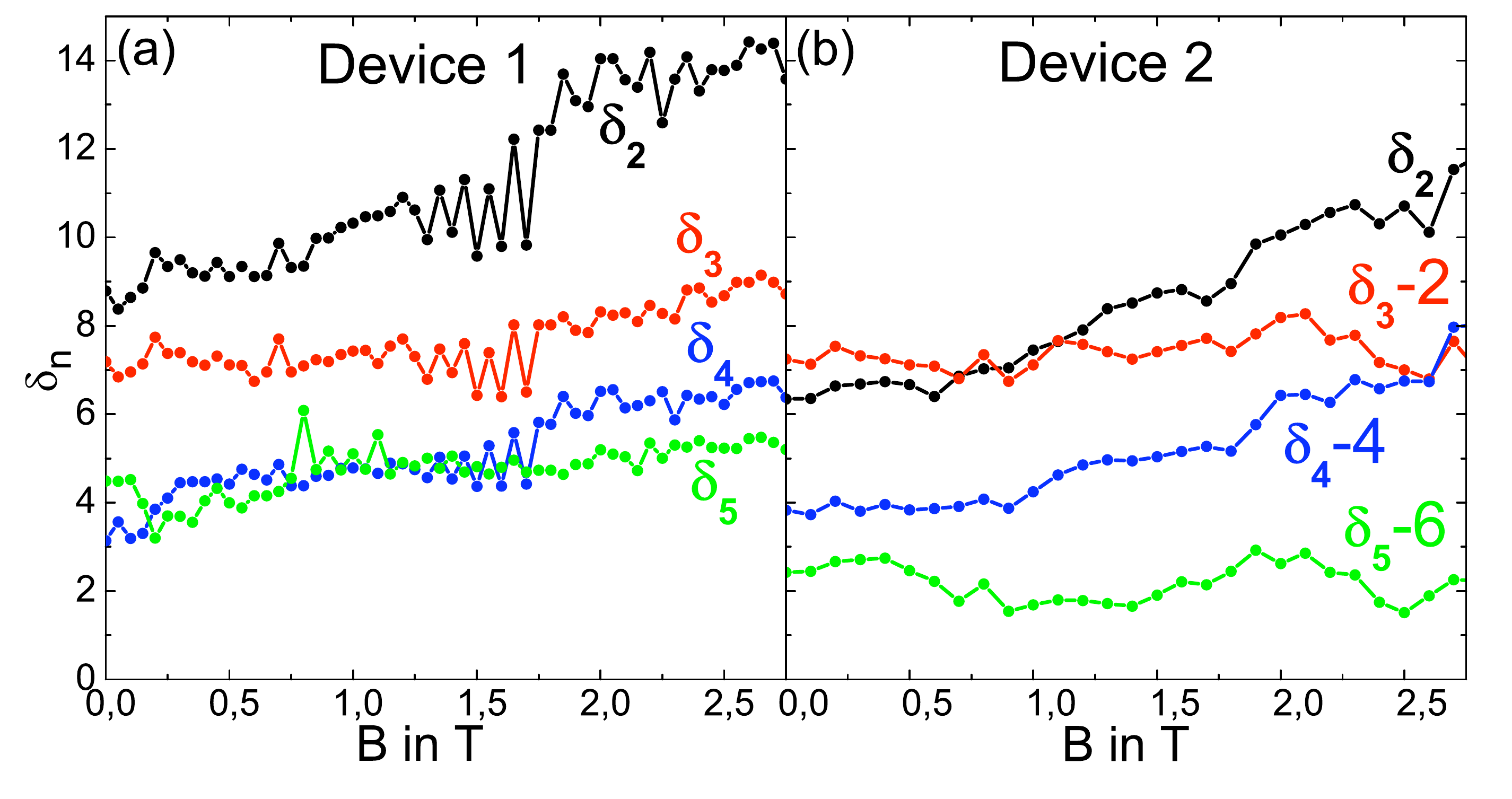}
  \caption{\label{fig:deltas} (Color online) $\delta_n$ as function of magnetic field strength of two different devices. Device 1 (a) was operated at a pumping frequency $f=50$ MHz and a power of $P=-16$ dBm. Device 2 (b) was operated at a $f=100$ MHz and $P=-11.5$ dBm, (b) is shifted for clarity.}
\end{figure}

Changes in the current plateaus caused by a perpendicular magnetic field have already been reported \cite{kaestner2009, wright2008}. From the investigations above we are able to visualize the changes separately for every plateau as a function of magnetic field. 

In Fig.~\ref{fig:deltas} measurement results are shown. The voltage $V^\text{DC}_L$ has been set to the middle of the plateau so that the vertical transition regions (corresponding to lines R$_+$ and L$_-$) have no siginificant effect on the current, i.e. insufficient filling or emptying of the QD is avoided. Due to the different carrier densities of the two devices, the required power for quantized pumping varies. For device 1 the power was set to $P=-16$ dBm and for device 2 to $P=-11.5$ dBm.
The graph shows $\delta_n$ obtained by fitting the normalized current to Eq.~\ref{eq:fit} as a function of the magnetic field. Despite the difference in the wafer characteristics, both devices show a reproducible even-odd behavior of the slopes. For even $n$ the corresponding $\delta_n$'s grow significantly faster with magnetic field than for odd $n$.

Device 1 shows for the first plateau an increase from $\delta_2(B=0)= 8.6$ to $\delta_2(B = 2.75\text{T})=13.2$. This means an increase of the tunnel coupling ratio $\Gamma^l_n/\Gamma^l_{n-1}$ by a factor of circa 100. Examining the third plateau the increase of $\Gamma^l_n/\Gamma^l_{n-1}$ is not as pronounced, but still raised by a factor of 15, due to the increase from $\delta_4(B=0)= 3.4$ to $\delta_4(B=2.75\text{T})= 8.6$. The second and fourth plateau show only a marginal increase. The tunnel coupling ratio is only raised by a factor of about 2.3 and 3, respectively. Device 2 reveals a similar result. While $\Gamma^l_n/\Gamma^l_{n-1}$ increases with magnetic field by a factor of 200 for the first and 65 for the third plateau, for the second and fourth plateau there is no significant increase visible. According to Ref.~\cite{kashcheyevs2009} the increase of the first plateau of both devices corresponds to an enhancement of the accuracy by two orders of magnitude. 

There are several factors contributing to the observed behavior. Energetic shifts should affect the decay rate ratios. Within a single-particle mean-field approach, an even-odd asymmetry in the field dependence of decay rate ratios may indicate sequential filling of orbitals. Ellipticity of the confining potential as well as deviations from constant-interaction approximation may complicate the B-field dependence of orbital energies \cite{kouwenhoven2001}.
Also the magnetic field dependent  barrier shape has to be kept in mind, due to its influence on screening. As a next step a microscopic theory of the tunnel couplings has to be found in order to obtain an exact information about the origin of this even-odd dependent evolution. This insight should be valuable to find strategies for systematic accuracy improvements for this pumping scheme.

\section*{Acknowledgements}
The research conducted within this EURAMET JRP "REUNIAM" has received funding from the European Community's Seventh Framework Programme, ERANET Plus, under Grant Agreement No.~217257. C.L. has been supported by International Graduate School of Metrology, Braunschweig. Assistance with device fabrication from P. Hinze and H. Marx and with the presentation from A.~Mueller are gratefully acknowledged.



\begin{thebibliography}{00}

  \bibitem{averin1986}
  D. V. Averin, K. K. Lihkarev, J. Low Temp. Phys. \textbf{62}, 345 (1986).
  
  \bibitem{mills2006}
  I. M. Mills, P. J. Mohr, T. J. Quinn, B. N. Taylor, E. R. Williams, Metrologia \textbf{43}, 227 (2006).

  \bibitem{nishiguchi2006}
  K. Nishiguchi, A. Fujiwara, Y. Ono, H. Inokawa, Y. Takahashi, Appl. Phys. Lett. \textbf{88}, 183101 (2006).

  \bibitem{barnes2000}
  C. H. W. Barnes, J. M. Shilton, A. M. Robinson, Phys. Rev. B \textbf{62}, 8410 (2000).

  \bibitem{geerligs1990}
  L. J. Geerligs, V. F. Anderegg, P. A. M. Holweg, J. E. Mooij, H. Pothier, D. Esteve, C. Urbina, M. H. Devoret, Phys. Rev. Lett. \textbf{64}, 2691 (1990).

  \bibitem{kouwenhoven1991}
  L. P. Kouwenhoven, A. T. Johnson, N. C. van der Vaart, C. J. P. M. Harmans, C. T. Foxon, Phys. Rev. Lett. \textbf{67}, 1626 (1991).

  \bibitem{pothier1992}
  H. Pothier, P. Lafarge, C. Urbina, D. Esteve, M. H. Devoret, Europhys. Lett. \textbf{17}, 249 (1992).
 
  \bibitem{pekola2008}
  J. P. Pekola, J. J. Vartiainen, M. Mšttšnen, O.-P. Saira, M. Meschke, D. V. Averin, Nat. Phys. \textbf{4}, 120 (2008).

  \bibitem{blumenthal2007}
  M. D. Blumenthal, B. Kaestner, L. Li, S. Giblin, T. J. B. M. Janssen, M. Pepper, D. Anderson, G. Jones, D. A. Ritchie, Nat. Phys. \textbf{3}, 343 (2007).

  \bibitem{kaestner2008}
  B. Kaestner, V. Kashcheyevs, S. Amakawa, M. D. Blumenthal, L. Li, T. J. B. M. Janssen, G. Hein, T. Weimann, U. Siegner, H. W. Schumacher, Phys. Rev. B \textbf{77}, 153301 (2008).

  \bibitem{fujiwara2008}
  A. Fujiwara, K. Nishigucchi, Y. Ono, Appl. Phys. Lett. \textbf{92}, 042102 (2008).

  \bibitem{kaestner2008b}
  B. Kaestner, V. Kashcheyevs, G. Hein, K. Pierz, U. Siegner, H. W. Schumacher, Appl. Phys. Lett. \textbf{92}, 192106 (2008).
  
  \bibitem{maire2008}
  N. Maire, F. Hohls, B. Kaestner, K. Pierz, H. W. Schumacher, R. J. Haug, Appl. Phys. Lett. \textbf{92}, 082112 (2008).

  \bibitem{kaestner2009}
  B. Kaestner, C. Leicht, V. Kashcheyevs, K. Pierz, U. Siegner, H. W. Schumacher, Appl. Phys. Lett. \textbf{94}, 012106 (2009).

  \bibitem{wright2008}
  S. J. Wright, M. D. Blumenthal, G. Gumbs, A. L. Thorn, M. Pepper, T. J. B. M. Janssen, S. N. Holmes, D. Anderson, G. A. C. Jones, C. A. Nicoll, D. A. Ritchie, Phys. Rev. B \textbf{78}, 233311 (2008).

  \bibitem{kashcheyevs2009}
  V. Kashcheyevs and B. Kaestner, cond-mat arXiv: 0901.4102.

  \bibitem{zole2009}
  Ritmars Zole, Bachelor Thesis, University of Latvia (2009).
 
  \bibitem{buettiker1990}
  M. B\"uttiker, Phys. Rev. B \textbf{41}, R7906 (1990).
 
  \bibitem{kouwenhoven2001}
  L. P.  Kouwenhoven, D. G. Austing, S. Tarucha, Rep. Prog. Phys. \textbf{64}, 701 (2001).



\end{thebibliography}
\end{document}